\documentclass[review,times,3p]{elsarticle}
\usepackage{graphicx}
\usepackage{amssymb}
\usepackage{amsthm}
\usepackage{amsmath}

\usepackage{lineno}

\journal{Nuclear Instruments and Methods in Physics Research Section A}

\newcommand{\code}[1]{{\ttfamily#1}}

\begin{document}
\linenumbers
\begin{frontmatter}

%% Title, authors and addresses

%% use the tnoteref command within \title for footnotes;
%% use the tnotetext command for theassociated footnote;
%% use the fnref command within \author or \address for footnotes;
%% use the fntext command for theassociated footnote;
%% use the corref command within \author for corresponding author footnotes;
%% use the cortext command for theassociated footnote;
%% use the ead command for the email address,
%% and the form \ead[url] for the home page:
%% \title{Title\tnoteref{label1}}
%% \tnotetext[label1]{}
%% \author{Name\corref{cor1}\fnref{label2}}
%% \ead{email address}
%% \ead[url]{home page}
%% \fntext[label2]{}
%% \cortext[cor1]{}
%% \address{Address\fnref{label3}}
%% \fntext[label3]{}

\title{A Novel Generic Framework for Track Fitting in Complex Detector Systems}

%% use optional labels to link authors explicitly to addresses:
%% \author[label1,label2]{}
%% \address[label1]{}
%% \address[label2]{}

\author{C. H\"oppner\footnote{corresponding author, email: christian.hoeppner@cern.ch}, S. Neubert, B. Ketzer, S. Paul}

\address{Technische Universit\"at M\"unchen, Physik Department, 85748 Garching, Germany}

\begin{abstract}
  This paper presents a novel framework for track fitting which is
  usable in a wide range of experiments, independent of the specific
  event topology, detector setup, or magnetic field arrangement. This
  goal is achieved through a completely modular design. Fitting
  algorithms are implemented as interchangeable modules. At present,
  the framework contains a validated Kalman filter. Track
  parameterizations and the routines required to extrapolate the track
  parameters and their covariance matrices through the experiment are
  also implemented as interchangeable modules. Different track
  parameterizations and extrapolation routines can be used
  simultaneously for fitting of the same physical track.
  Representations of detector hits are the third modular ingredient
  to the framework. The hit dimensionality and orientation of planar
  tracking detectors are not restricted. Tracking information from
  detectors which do not measure the passage of particles in a fixed
  physical detector plane, e.g.\ drift chambers or TPCs, is used
  without any simplifications. The concept is implemented in a
  light-weight C++ library called GENFIT, which is available as free
  software.
\end{abstract}

\begin{keyword}
%% keywords here, in the form: keyword \sep keyword
track fitting \sep track reconstruction \sep Kalman filter \sep drift
chamber \sep TPC
%% PACS codes here, in the form: \PACS code \sep code

%% MSC codes here, in the form: \MSC code \sep code
%% or \MSC[2008] code \sep code (2000 is the default)

\end{keyword}
\end{frontmatter}

%% \linenumbers

%% main text
\section{Introduction}
\label{sec:introduction}
Spectrometers in nuclear and particle physics have the purpose of
identifying the 4-momenta and vertices of particles stemming from
high-energy collisions and decays of particles or nuclei. In addition
to calorimetric and other particle identification measurements, the
3-momenta and positions of charged particles are measured by tracking
them in magnetic fields with the use of position sensitive
detectors. Cluster finding procedures can be applied in some detectors
to combine the responses of individual electronic channels in order to
improve the accuracy of the position measurements. The position
measurements will be called \emph{hits} throughout this paper,
regardless of whether they stem from a single detector channel or from
a combination of several of them. Pattern recognition algorithms
determine which hits contribute to the individual particle tracks.
The hits identified at this stage to belong to one track then serve as
the input for a fitting procedure, which determines the best estimates
for the position and momentum of a particle at any point along its
trajectory. A novel framework for this task of track fitting in complex detector
systems is presented in this paper. It organizes the
  task of track fitting, i.e.\ the interplay between fitting algorithms,
  detector hits, and particles trajectories, with a minimal amount of
  interfaces. It is a toolkit which is
independent of specific detector setups and magnetic field geometries
and hence can be used for many particle physics
experiments.

Tracking of particles is usually performed with a combination of
different species of detectors. They can be categorized according to
the different
geometrical information they deliver:\\
%\begin{itemize}
1) detectors which measure the particle passage along one axis in a
detector plane, e.g.\ silicon strip detectors or multiwire
proportional chambers;\\ 2) detectors which measure the
two-dimensional penetration point of a particle through a plane, e.g.\
silicon pixel detectors;\\ 3) detectors which measure a drift time
relative to a wire position, i.e.\ a surface of constant drift time
around the wire through which the particle passed tangentially, e.g.\
drift chambers or ``straw tubes'';\\ 4) detectors which measure
three-dimensional space points on particle trajectories, like time
projection chambers (TPC).\\ But also higher dimensional hits can
occur:\\ 5) detector systems which measure two-dimensional position
information in combination with two-dimensional direction information,
including correlations between these parameters. Examples could be
stations of several planes of detectors of categories 1 and 2, or
electromagnetic calorimeters.\\For detectors which do not deliver
tracking information in physical detector planes, e.g.\ those of
categories 3 and 4, the track fitting software of many experiments
resorts to simplifications, which may be justified for a particular
application but prevent the usage of the same program for different
experimental environments. Examples are the projection of TPC data
onto planes defined by pad rows or the projection of the surfaces of
constant drift time in drift chambers onto predefined planes, just
leaving two lines with left-right ambiguities. This approach is
problematic if the drift cells are not arranged in a planar
configuration and if there is no preferred direction in which the
detector is passed by the particles. Another common simplification is
the treatment of two-dimensional hits (e.g.\ from silicon pixel
detectors) as two independent one-dimensional measurements.\\ In the
framework presented here these problems have been overcome to make
optimal use of the information from combinations of all types of
tracking detector systems.  All detector hits are defined in detector
planes. For hits in detectors which do not have physical detector
planes, so-called \emph{virtual detector planes} are calculated
dynamically for every extrapolation of a track to a hit. The
dimensionality of detector hits is not restricted. One-dimensional
hits constrain the track only along the coordinate axis in the
detector plane which they measure. Two-dimensional hits are used in
one fitting step to constrain the track in two dimensions in their
detector planes.  For hits in non-planar detectors (categories 3 and
4), the hit information (e.g.\ a surface of constant drift time) is
converted into a position measurement in a plane perpendicular to the
track, so that a fit is able to minimize the perpendicular distances
between the track and the position measurements.  The information from
hits with higher dimensionality, like those of category 5, is used in
four-dimensional hits, which contain all correlations between the
parameters.\\ Tracks of charged particles in magnetic fields are
(usually) described by five parameters and a corresponding covariance
matrix. The ability to extrapolate a track described by these
parameters and their covariances, taking into account the effects of
materials and magnetic fields, to different positions in the
spectrometer is mandatory for track fitting.  The concept presented
here provides a well defined interface for the invocation of external
programs or libraries to perform these track extrapolations. It
  thus allows the straightforward use of established track following
  codes with their native geometry and magnetic field interfaces, such
  as GEANE \cite{GEANE}, which is nowadays distributed as part of
  CERN's Virtual Monte Carlo (VMC) package \cite{VMC}. This is the
  most significant difference to other projects (e.g.\ RecPack
  \cite{RECPACK}), which offer more monolithic approaches to track
  fitting (e.g.\ defining their own geometry classes). The concept
allows the simultaneous fitting of several representations of tracks
to the same set of hits, i.e.\ to the same physical track. This
flexibility is especially useful in the early phase of an experiment
when different track parameterizations and extrapolation approaches
can be compared with each other, in order to identify the ones with
optimal performance. But also the flexible coverage of different phase
space regions with different track models, or the fitting of different
mass hypotheses with the same track model can be desirable. The
implementation of the concept has been realized in a software toolkit
called GENFIT. It is written in C++ and is designed in a fully object
oriented way. It has been developed in the framework of the PANDA
experiment \cite{PANDA}, as part of the computing framework PANDAroot
\cite{PANDAroot}, but is now distributed as a stand-alone package
\cite{genfitSF}.\\ GENFIT contains a validated Kalman Filter. This
algorithm is commonly used for track fitting in particle spectrometers
\cite{KalmanFit}, since it performs much better than global
minimization approaches in the presence of materials and inhomogeneous
magnetic fields. The concept is however not limited to the use of the
Kalman Filter. Other fitting algorithms, like Gaussian Sum Filters
\cite{gaussSum} or Deterministic Annealing Filters \cite{daf}, can be
implemented easily.

Section \ref{sec:concept} describes the concept of this new approach
to track fitting in detail. Section \ref{sec:implementation} points
out the key features of the implementation of GENFIT. Some examples of
concrete track representations, on
the dimensionalities of reconstruction hits and track representations,
and the interplay between them follow in Sec.\ \ref{sec:examples}. Simulation studies which validate the Kalman
filter implemented in GENFIT are presented in
Sec.\ \ref{sec:simulation}.

\section{Concept}     
\label{sec:concept}
\begin{figure*}[htb]
  \begin{center}
    \includegraphics[width=0.99\textwidth]{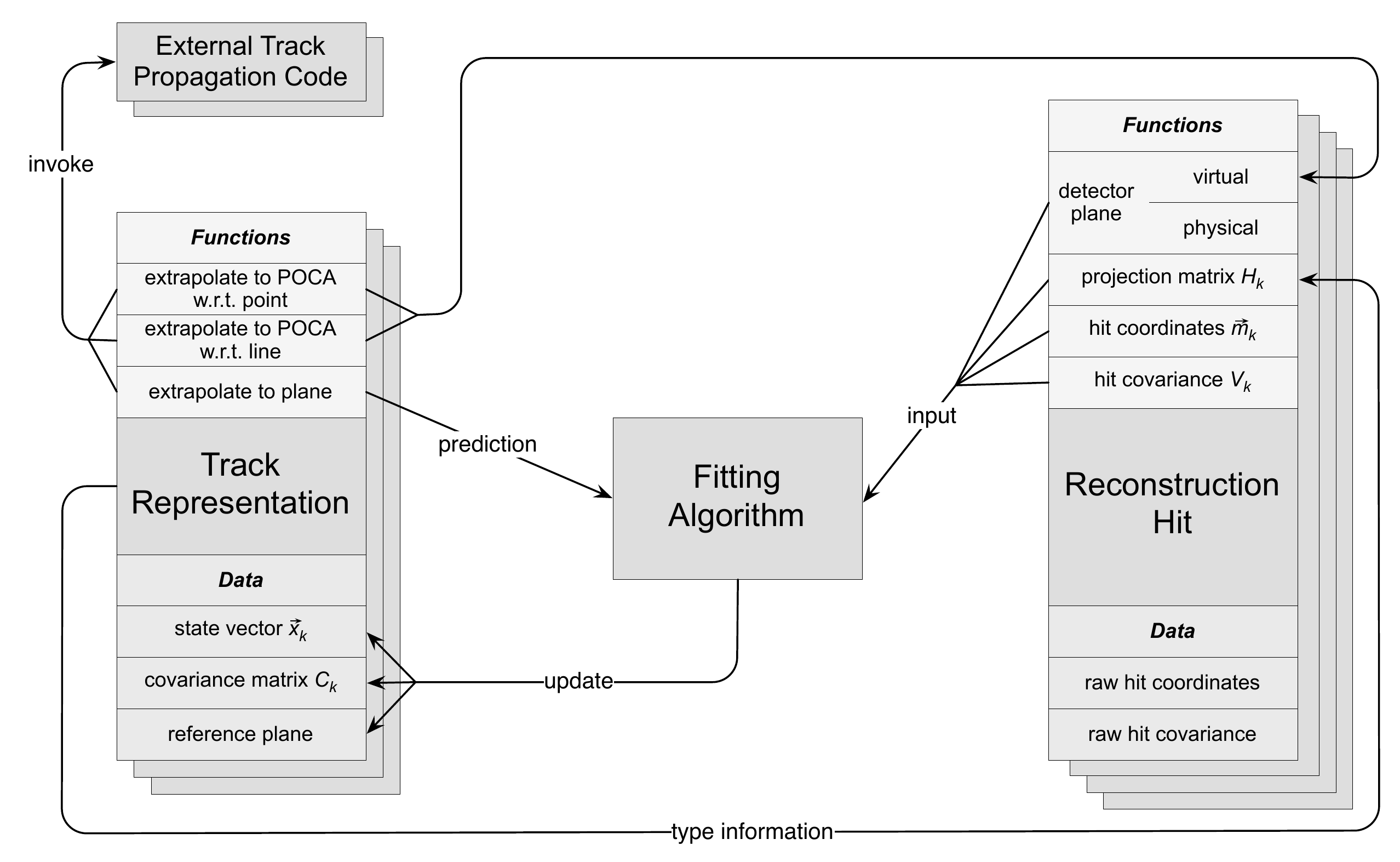}
    \caption{General structure of objects for track fitting: Fitting
      Algorithm, Track Representation, and Reconstruction Hit. The
      arrows indicate the interactions between the objects, which are
      described in this chapter. POCA stands for point of closest
      approach.}
    \label{fig:overview}
  \end{center}
\end{figure*}

The basic functionalities which are required for any procedure of track
fitting are the extrapolation of tracks to the positions of the hits
in the detectors, and the calculation of the distances between hits
and tracks, i.e. the residuals. The concept discussed here divides the
problem of track fitting into three main entities which are separated
from each other as much as possible and interact through well defined
interfaces: 1) track fitting algorithms, 2) track representations, and
3) reconstruction hits. Figure \ref{fig:overview} shows this
structure. The following sections explain these objects in detail.
\subsection{Track Fitting Algorithms}
\label{sec:concept:algorithm}
``Progressive'' fitting algorithms like the extended Kalman filter \cite{KalmanFit,Kalman} are widely used for track fitting
in high energy physics experiments.
Although the track fitting concept discussed
in this paper is not limited to the use of the Kalman filter, this
algorithm shall serve as an example to illustrate which
functionalities are generally required.\\ The extended Kalman filter
is an efficient recursive algorithm that finds the optimum estimate
$\vec{x}_k$ for the unknown true state vector $\hat{\vec{x}}_k$ of a
system from a series of noisy measurements, together with the
corresponding covariance matrix $C_k$ of $\vec{x}_k$. The state vector
contains the track parameters and the index $k$ indicates that the
state vector, and its covariance matrix are given at the detector
plane of hit $k$.\\ Before a recursion step, the state vector
$\vec{x}_{k-1}$ and covariance matrix $C_{k-1}$ contain the
information of all hits up to index $k-1$. In the \textbf{prediction}
step the state vector and covariance matrix are extrapolated to the
detector plane of hit $k$ by the track following code. The predicted
state vector is denoted by $\tilde{\vec{x}}_k$ and the predicted
covariance matrix by $\tilde{C}_k$.  This covariance matrix is the sum
of the propagated track covariance matrix $C_{k-1}$ (Gaussian error
propagation by transformation with the Jacobian matrix of the
propagation operation $\tilde{\vec{x}}_k=f(\vec{x}_{k-1}))$, and a
noise matrix which takes into account effects like multiple
scattering and energy loss straggling.  Then, the algorithm calculates
the \textbf{update} for the state vector and the covariance matrix by
taking into account the measurement $\vec{m}_k$:
\begin{align}
  &\vec{x}_{k}=\tilde{\vec{x}}_{k}+K_{k}\tilde{\vec{r}}_{k}\\
  &C_{k}=(I-K_kH_k)\tilde{C}_{k}
\end{align}
with the residual
\begin{align}
 &\tilde{\vec{r}}_{k} = \vec{m}_k - H_k\tilde{\vec{x}}_{k},\label{eq:resid}
\end{align}
the weight of the residual (or Kalman gain)
\begin{align}
&K_{k} = \tilde{C}_{k}H_k^T(H_k\tilde{C}_{k}H_k^T+V_k)^{-1},
\end{align}
and the covariance matrix $V_k$ of the measurement $\vec{m}_{k}$.  $I$
is the unit matrix of corresponding dimensionality. The projection
matrix $H_k$ is a linear transformation from the coordinate system of
the state vector $\vec{x}_k$, to the coordinate system of the position
measurement $\vec{m}_k$ of hit $k$, i.e.\ the detector plane of the
hit. A discussion about dimensions of the vectors and matrices in the
above equations can be found in Sec.\ \ref{sec:examples:interplay}
together with concrete examples for the matrix $H_k$. The elements of
the covariance matrix $C_k$ shrink with the inclusion of more hits,
thus reducing the impact of a single hit on the value of the state
vector. The $\chi^2$-contribution of hit $k$ is $\chi^2_k =
\vec{r}_{k}^{\ T}(V_k-H_kC_{k} H_k^T)^{-1}\vec{r}_{k}$ with the
filtered residual $\vec{r}_{k} = \vec{m}_k - H_k\vec{x}_{k}$. It adds
$\text{dim}(\vec{m}_k)$ degrees of freedom to the total $\chi^2$.\\
After the Kalman steps have been performed on all hits of the track,
the track can still be biased due to wrong starting values
$\vec{x}_{0}$. This bias can be reduced by the repeated application of
the procedure in the opposite order of hits, using the previous fit
result as starting values for the track parameters. Before the
  fit is repeated, the elements of the covariance matrix have to be
  multiplied with a large factor ($\mathcal{O}(1000)$) in order not to
  include the same information in the track several times.\\ As can
be seen in Fig.\ \ref{fig:overview} the fitting algorithm operates on
entities called reconstruction hits and track representations, which
are detailed in the following.
\subsection{Track Representations}
\label{sec:concept:trackrep}
A particle track is described by a set of track parameters and a
corresponding covariance matrix, which are defined at a given position
along the track. Often, the track parameters are e.g.\ given at a
particular $z$-position. In the concept presented here, track
parameters are always defined in reference planes.\\ In order to use a
track model in a track fitter, one needs to be able to extrapolate the
track parameters to different places in the spectrometer. The
combination of the track parameterization and the track extrapolation
functionality will be called a \emph{track representation}.  A track
representation holds the data of the state vector $\vec{x}_k$, and
the covariance matrix $C_k$ of a track, as well as the reference
plane at which these are defined. Also it provides a well defined
interface for the invocation of the external routines needed to
extrapolate the parameters to different positions.  As can be seen in
Fig.\ \ref{fig:overview}, there are three track extrapolation
functions which are needed for each track representation:
Extrapolation to a plane, extrapolation to the point of closest
approach (POCA) to a point, and extrapolation to the point of closest
approach to a line. Fitting algorithms access the track parameters and
extrapolation functions in a common way via the track representation
interface without knowledge of the specific form of the track
parameterization or the way the
extrapolations are carried out.\\
Different track representations can be used in parallel. It is
possible to fit the same track, i.e.\ the same set of hits, with
different track representations simultaneously. There are several
reasons why this is desirable: For low momentum particles the fitting
of different mass hypotheses with the same track representation can
give a clue to the particle identity via the $\chi^2$ of the fits,
because the different energy loss for different particle masses at a
given momentum leads to different extrapolations. Fitting of the same
track with different parameterizations and extrapolation tools can be
advantageous as well. In the early phase of an experiment one can
compare different track representations to identify the ones which
perform best, or there could be regions in phase space in experiments
where it might not be clear beforehand which track representation will
give the best results. Then one can just fit several of them
simultaneously and retain the best result.
\subsection{Reconstruction Hits}
\label{sec:concept:recohit}
\begin{figure}[t]
  \begin{center}
    \includegraphics[width=\columnwidth]{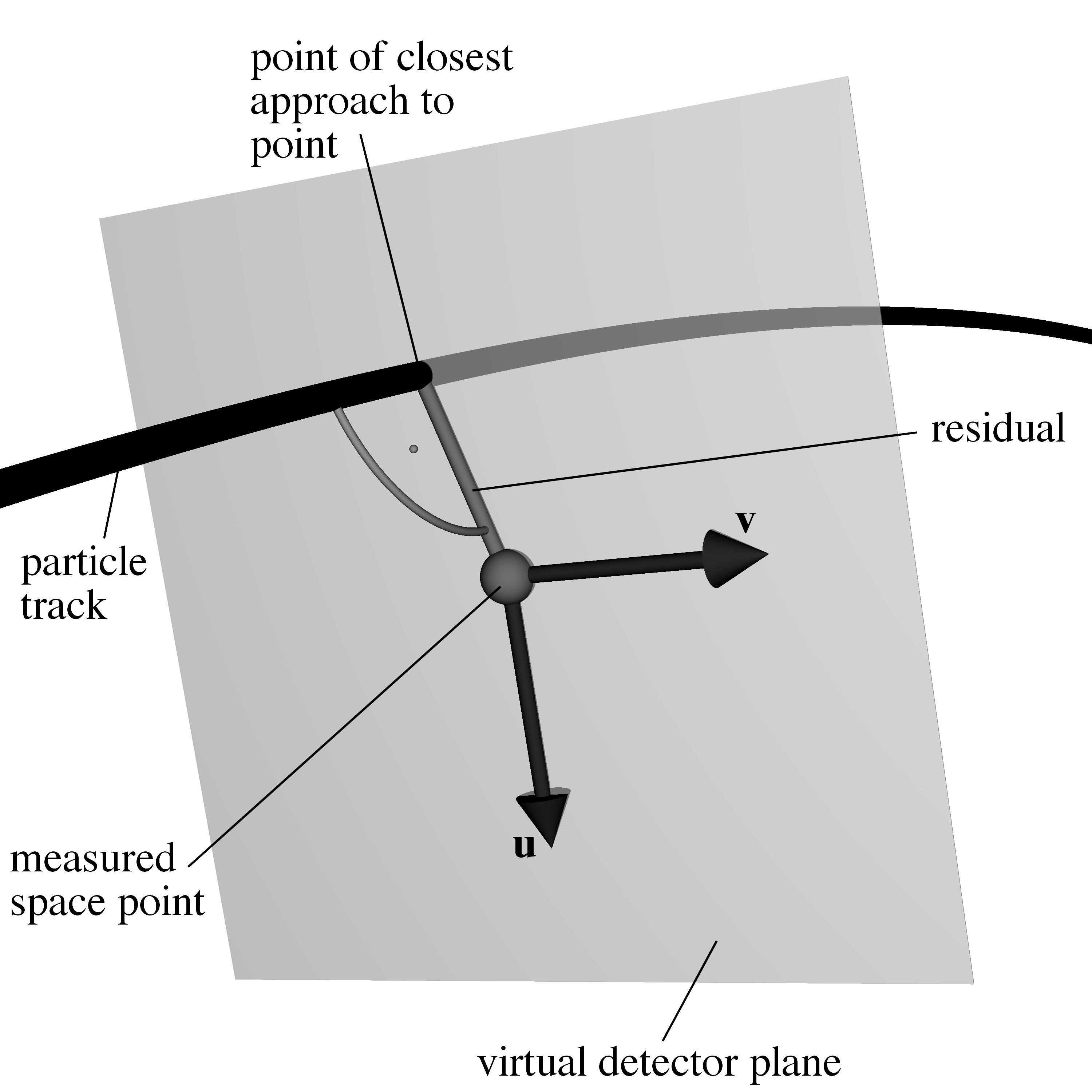}
    \caption{Virtual
      detector plane (spanning vectors $\vec{u}$ and $\vec{v}$) for a space-point hit.}
    \label{fig:genfit:virt}
  \end{center}
\end{figure}

\begin{figure}[t]
  \begin{center}
    \includegraphics[width=\columnwidth]{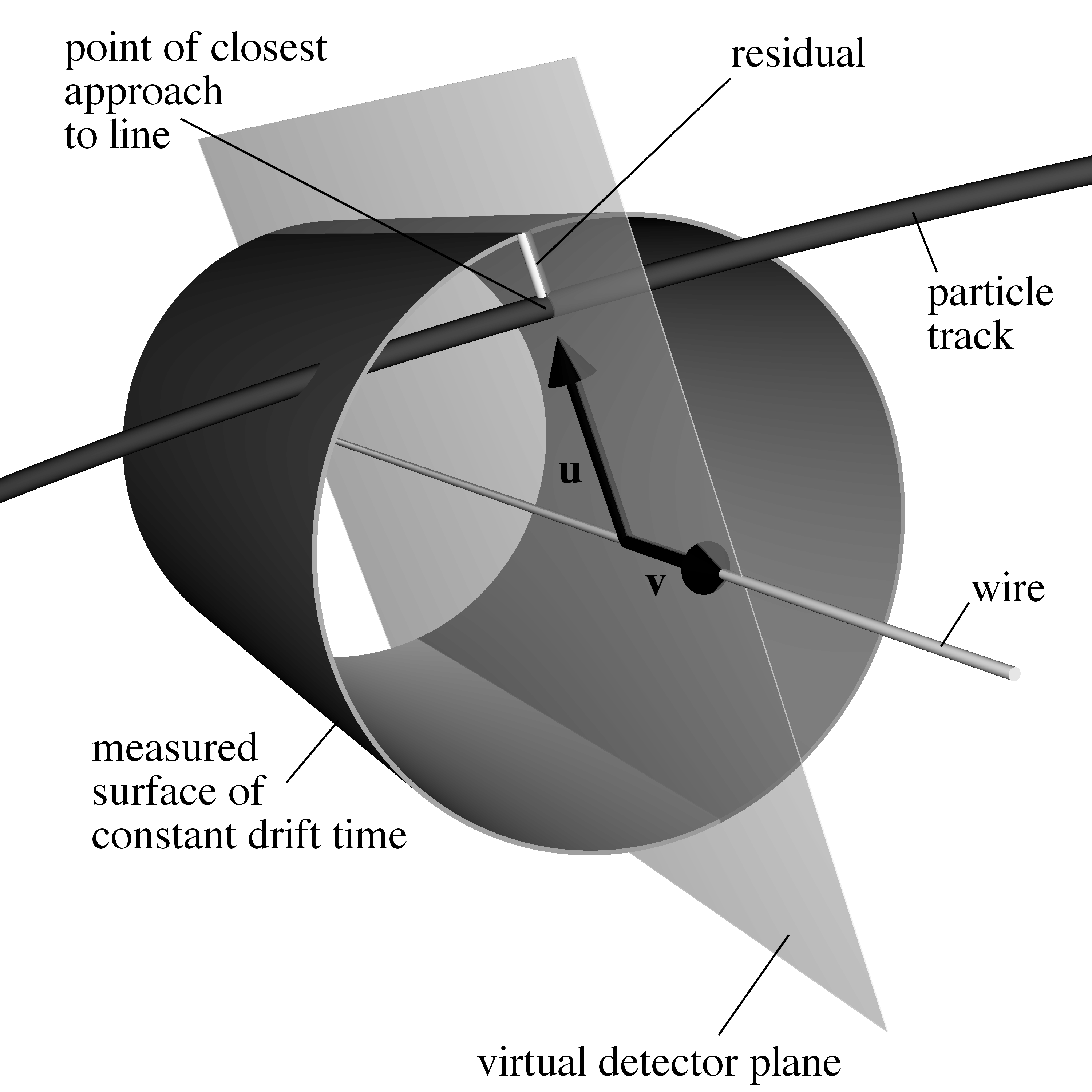}
    \caption{Virtual detector plane (spanning vectors $\vec{u}$ and $\vec{v}$) for a wire-based drift detector.}
    \label{fig:genfit:virtSTT}
  \end{center}
\end{figure}

The object which represents a position measurement from a detector
used in a track fit is called a \emph{reconstruction hit}. It contains
the vector of the raw measurement coordinates and its corresponding
covariance matrix. As discussed in the introduction, the nature of
this raw hit information can be quite diverse. It can e.g.\ be a
direct position measurement or a drift time. As can be seen in Fig.\
\ref{fig:overview}, a reconstruction hit provides its detector
plane, the measurement coordinates $\vec{m}_k$ in
the detector plane coordinate system, the covariance matrix $V_k$ in
the detector plane coordinate system, and the projection matrix
$H_k$ to the fitting algorithm. For detectors, which measure positions in a physical detector
plane (categories 1 and 2 of Sec.\ \ref{sec:introduction}), the
detector plane is identical with the physical plane.\\
For non-planar detectors like wire-based drift chambers or TPCs
  (categories 3 and 4 of Sec.\ \ref{sec:introduction}), no such
  physical detector planes are defined. Instead, the concept of
  virtual detector planes is introduced. For space-point detectors,
  the track fit has to minimize the perpendicular distances of the
  track to the hits. Therefore, the virtual detector plane for each
  hit must contain the hit position and the point of closest approach
  of the track to the hit point. Then the residual vector which points
  from the hit point to the point of closest approach will be
  perpendicular to the track. This geometry is illustrated in Fig.\
  \ref{fig:genfit:virt}. The orientation of the spanning vectors
  $\vec{u}$ and $\vec{v}$ is chosen arbitrarily in the plane. For
  wire-based drift detectors the virtual detector plane contains the
  point of closest approach of the track to the wire, and is oriented
  to contain the whole wire. The spanning vectors are chosen to lie
  perpendicular ($\vec{u}$) and along ($\vec{v}$) the wire. This
  geometry is shown in Fig.\ \ref{fig:genfit:virtSTT}. The wire
  position and drift time are then measurements of $u$ (the $v$
  coordinate could be measured via double-sided readout with charge
  sharing or time of propagation). In both cases, the orientation of
  the plane will directly depend on the track parameters. The
  consequence is that virtual detector planes have to be calculated
  each time a hit is to be used in a fitting step. The reconstruction
  hit uses the corresponding extrapolation function of the given track
  representation to find the point of closest approach as indicated in Fig.\ \ref{fig:overview}.\\
Different kinds of reconstruction hits are accessed via a common
interface. When the fitting algorithm obtains the detector plane from
a reconstruction hit, it does not know whether it will receive a
physical or a virtual detector plane. This distinction is fully
handled inside the reconstruction hit.\\
After the detector plane is defined, the reconstruction hit can
provide the measurement coordinate vector $\vec{m}_k$, and the hit
covariance matrix $V_k$. For non-planar detectors, these quantities
are results of coordinate transformations into the virtual detector
plane (hence the difference between the raw hit
coordinates/covariance and the vector $\vec{m}_k$ and matrix $V_k$ in
Fig.\  \ref{fig:overview}). The three-dimensional hit vector and the
$3\times 3$ covariance matrix of a space-point hit are transformed
into a two-dimensional vector in the detector plane and a $2\times 2$
covariance matrix. Even if the errors of the space point were
uncorrelated, the matrix $V_k$ will in general contain a correlation,
which is taken into account in the fit. For wire-based drift chambers,
the drift time information is converted to
a position information in the calculation of $\vec{m}_k$ and $V_k$.\\
The projection matrix $H_k$ transforms the state vector from the
given track parameterization into the coordinate system of the hit. In
order to determine this matrix, the concrete coordinate systems of the
track representation and the reconstruction hit must be known. Since there
will be typically more different types of reconstruction hits than
track representations, the projection matrix is determined in the
reconstruction hit object. The
matrix $H_k$ provides the only link between a given track
parameterization and the different hit coordinate systems. If a fit is
performed with several track representations, the same reconstruction
hit will provide a different matrix $H_k$ for each track
representation.

\section{Implementation - GENFIT}
\begin{figure*}[htb]
  \begin{center}
    \includegraphics[width=.9\textwidth]{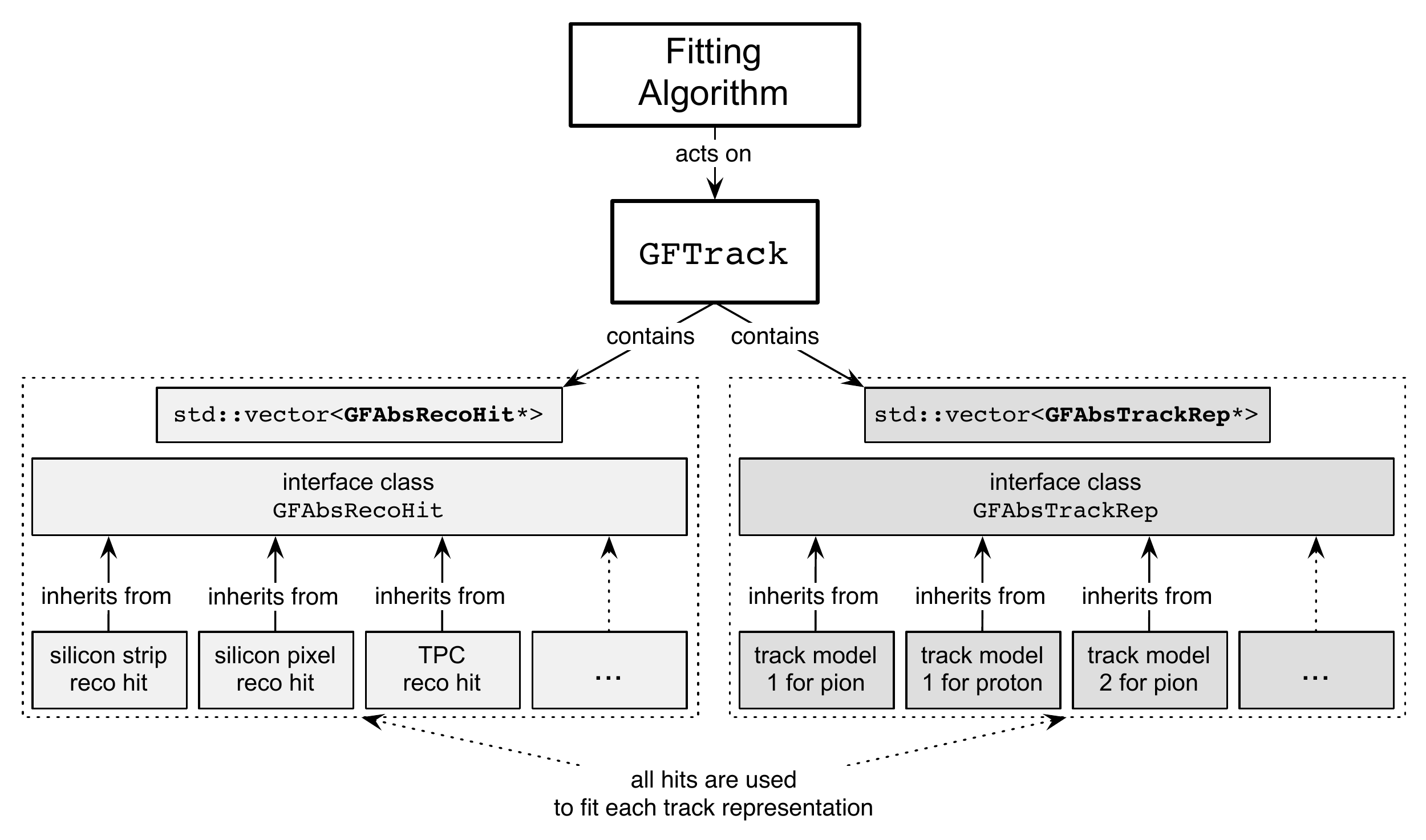}
    \caption{Class structure of GENFIT. The detailed inheritance
      structure of reconstruction hits is shown in Fig.\ \ref{fig:genfit:recohit}.}
    \label{fig:genfit:general}
  \end{center}
\end{figure*}
The software package which implements the concept presented in this
paper is called GENFIT \cite{genfitSF}. It is completely written in
C++ and makes extensive use of object oriented design. It uses the C++
standard template library \cite{Stroustrup} and the ROOT data analysis
framework \cite{ROOT}.\\ Figure \ref{fig:genfit:general} presents the
general class structure of GENFIT. The classes representing the
fitting algorithms operate on instances of the class
\code{GFTrack}\footnote{class names or other code fragments are set in
  \code{typewriter font}.}. A \code{GFTrack} object contains a
\code{std::vector<GFAbsRecoHit*>} and a\\
\code{std::vector<GFAbsTrackRep*>}. The reconstruction hits and track
representations of Secs.\ \ref{sec:concept:trackrep} and
\ref{sec:concept:recohit} are realized as polymorphic classes. The
class \code{GFAbsRecoHit} is the interface class to the reconstruction hits,
and \code{GFAbsTrackRep} is the interface class to the track
representations.\\ The reconstruction hit objects are created from the
position information acquired in the detectors. The pattern
recognition algorithms, which precede the use of GENFIT, determine
which of these detector hits belong to a certain track. They deliver
an instance of the class \code{GFTrackCand}, which holds a list of
indices which identify the hits belonging to the track. A mechanism
called \code{GFRecoHitFactory} has been implemented to load the
reconstruction hits into the \code{GFTrack} object.

\label{sec:implementation}

\subsection{Track Representations}
\label{sec:implementation:trackrep}
In order to use a particular track parameterization for track fitting
in GENFIT, one needs code which can extrapolate such track
parameters, taking into account material effects on the track
parameters and their covariance matrix. In order to interface
the track model to GENFIT, one implements a C++ class which inherits
from the abstract base class \code{GFAbsTrackRep} and provides an
implementation for the virtual methods
\code{extrapolate(...)}, \code{extrapolateToPoint(...)}, and
\code{extrapolateToLine(...)}. Section
\ref{sec:examples:geanetrackrep} presents examples of concrete
track representations.
\subsection{Reconstruction Hits}
\label{sec:implementation:recohits}
\begin{figure*}[htb]
  \begin{center}
    \includegraphics[width=.99\textwidth]{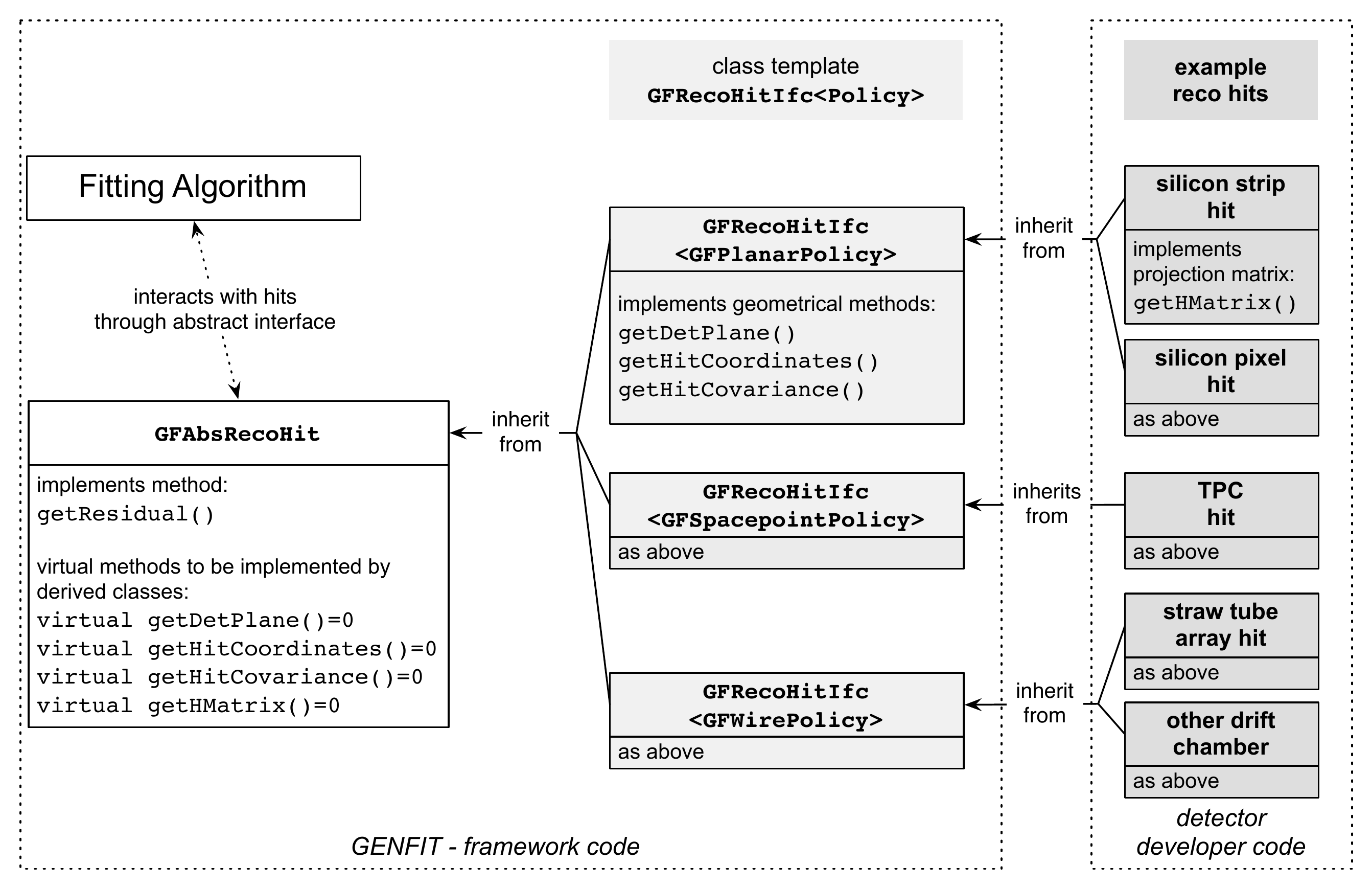}
    \caption{Inheritance structure of reconstruction hits in GENFIT.}
    \label{fig:genfit:recohit}
  \end{center}
\end{figure*}
The fitting algorithms interact with the reconstruction hits via the
abstract base class \code{GFAbsRecoHit}. The reconstruction hits do,
however, not inherit directly from this class, but from the
intermediate interface class\\ \code{GFRecoHitIfc<Policy>}. This is
illustrated in Fig.\ \ref{fig:genfit:recohit}. For more information
about the policy design pattern, please see \cite{alexandrescu}. There
are (currently) three geometrical categories of reconstruction hits:
Hits in planar detectors, space-point hits, and hits in wire-based
drift chambers, which deliver their wire position and a drift
time. This categorization is expressed in the code by the three
different policy classes \code{GFPlanarPolicy},
\code{GFSpacepointPolicy}, and \code{GFWirePolicy}. These policy
classes all implement functions for calculating or delivering the
detector plane, the hit coordinates in the detector plane, and the hit
covariance matrix in the detector plane. They are used to unify the
geometrical properties of reconstruction hits to avoid any code
duplication in the implementation of similar reconstruction hits. The
latter two policies use the corresponding track representations to
calculate the virtual detector planes, as detailed in Sec.\
\ref{sec:concept:recohit}.\\ As described in Sec.\
\ref{sec:concept:algorithm}, the fitting algorithm needs a matrix
$H_k$ which is a linear transformation from the vector space of track
parameters to the coordinate system defined by the detector plane. The
virtual method\\ \code{GFAbsRecoHit::getHMatrix(...)} is overridden in
the implementations of the concrete reconstruction hits.  In order to
provide the correct matrix, the reconstruction hit determines the
concrete type\footnote{by performing a C++ \code{dynamic\_cast} on the
  base class pointer \code{GFAbsTrackRep*}.} of the track
representation it is asked to interact with in this particular fitting
step. This type checking is represented by the lower arrow in Fig.\
\ref{fig:overview}. It is the only place in GENFIT where a
direct type compatibility of tracks and hits is checked. A maximal
modularity of the system is achieved through this mechanism. If one
adds an additional track representation, it is quite obvious that one
has to provide new coordinate transformations from this new parameter
space into the coordinate systems in which the hits are defined.

\section{Examples}
\label{sec:examples}
\subsection{Concrete Track Representations}
\label{sec:examples:geanetrackrep}
A concrete interface to an external track propagation package which
has been realized with GENFIT is the track representation called
\code{GeaneTrackRep2}. It is based on the FORTRAN code GEANE. The detector geometry is
included via the \code{TGeo} classes of ROOT \cite{ROOT} and the
magnetic field maps are accessed via a simple interface class called \code{GFAbsBField}. State
vectors for this track representation are defined as
$\vec{x}_k=(q/\vert \vec{p}\vert,du/dw,dv/dw,u,v)^T$, where the
detector plane is spanned by the vectors $\vec{u}$ and $\vec{v}$
(normal vector $\vec{w} = \vec{u} \times \vec{v}$). $q$ denotes the
particle charge and $\vec{p}$ is the particle momentum. The
quantitative tests of GENFIT in Sec.\ \ref{sec:simulation} are carried
out with this track representation.\\ Another track representation
included in the GENFIT distribution is called \code{RKTrackRep}. It
was adopted from the COMPASS experiment \cite{COMPASS} and uses a
Runge-Kutta solver to follow particles through magnetic fields. It has
the same state vector definition as \code{GeaneTrackRep2}. It also
uses the \code{TGeo} classes for the geometry interface.
\subsection{Interplay between Track Representations and Reconstruction
  Hits}
\label{sec:examples:interplay}
The classes which represent the fitting algorithms just carry out
their linear algebra without knowing about the dimensions of the state
vectors $\vec{x}_k$ and the measurement vectors $\vec{m}_k$.
The matrix $H_k$ is provided by the reconstruction hit class to
transform state vectors and covariance matrices of a specific parameterization into the
measurement vector coordinate system. This projection matrix ensures
that the dimensionalities of the vectors and matrices in the fitting algorithm are
compatible with each other. The following examples shall illustrate this:
\begin{enumerate}
\item A four-dimensional track model can be used for tracking without
  magnetic fields. The state vector is defined as
  $\vec{x}_k=(u,v,du/dw,dv/dw)^T$ for a straight line where $\vec{u}$
  and $\vec{v}$ span the detector plane, and $\vec{w}=\vec{u} \times
  \vec{v}$ is the normal vector. A strip detector shall measure the
  $u$ coordinate. Then the measurement vector of equation (\ref{eq:resid}),
  $\vec{m}_k$, is a scalar. The projection matrix is defined as
  $H_k=(1,0,0,0)$, so that $H_k\cdot \vec{x}_k$ is one-dimensional,
  just as the residual $\tilde{\vec{r}}_{k}$. The Kalman gain is a
  $4\times 1$ matrix, and the $\chi^2$-increment is correctly
  calculated for one degree of freedom, in the sense that
  $\vec{r}_{k}$ and $(V_k-H_k C_{k} H_k^T)$ are scalars.
\item A pixel detector is used in combination with a five-dimensional
  trajectory model for charged particle tracking in magnetic fields.
  The detector measures the coordinates $u$ and $v$ in the detector
  plane, and the state vector is $\vec{x}_k=(q/\vert
  \vec{p}\vert,du/dw,dv/dw,u,v)^T$. The $2\times5$ projection matrix
  is then:
\begin{align*}
  H_k =  \left( \begin{array}{ccccc}
      0 & 0 & 0 & 1 & 0\\
      0 & 0 & 0 & 0 & 1\end{array} \right)\
\end{align*}
All matrices and vectors automatically appear with correct
dimensions: $\vec{m}_k$ and $\tilde{\vec{r}}_{k}$ are 2-vectors, $V_k$ is a
$2\times2$ matrix, the Kalman gain is a $5\times2$ matrix, and
$\chi^2$ is a scalar which is calculated from two degrees of freedom
($\vec{r}_{k}$ is a 2-vector, and $(V_k-H_k C_{k} H_k^T)$ is a
$2\times 2$ matrix).\\ If the next hit in the same track only measures one
coordinate (e.g.\ $u$ in the detector plane coordinate system of the
next hit) , $\vec{m}_k$ will be scalar, $H_k$ will be of dimension
$1\times 5$, and there will be only one degree of freedom added to the
overall $\chi^2$.
\item A TPC delivers space-point hits. The track model is the same as
  in example 2. The TPC measures three spatial coordinates but this
  information is transformed into a two-dimensional hit in the virtual
  detector plane, which is perpendicular to the track. This
  two-dimensional hit is treated identically to example 2. This is the
  desired behaviour, since measurements or errors along the flight
  direction do not contribute to the track fit.
\end{enumerate}

\section{Simulation Studies}
\label{sec:simulation}
\begin{figure*}[p]
  \begin{center}
    \includegraphics[width=.99\textwidth]{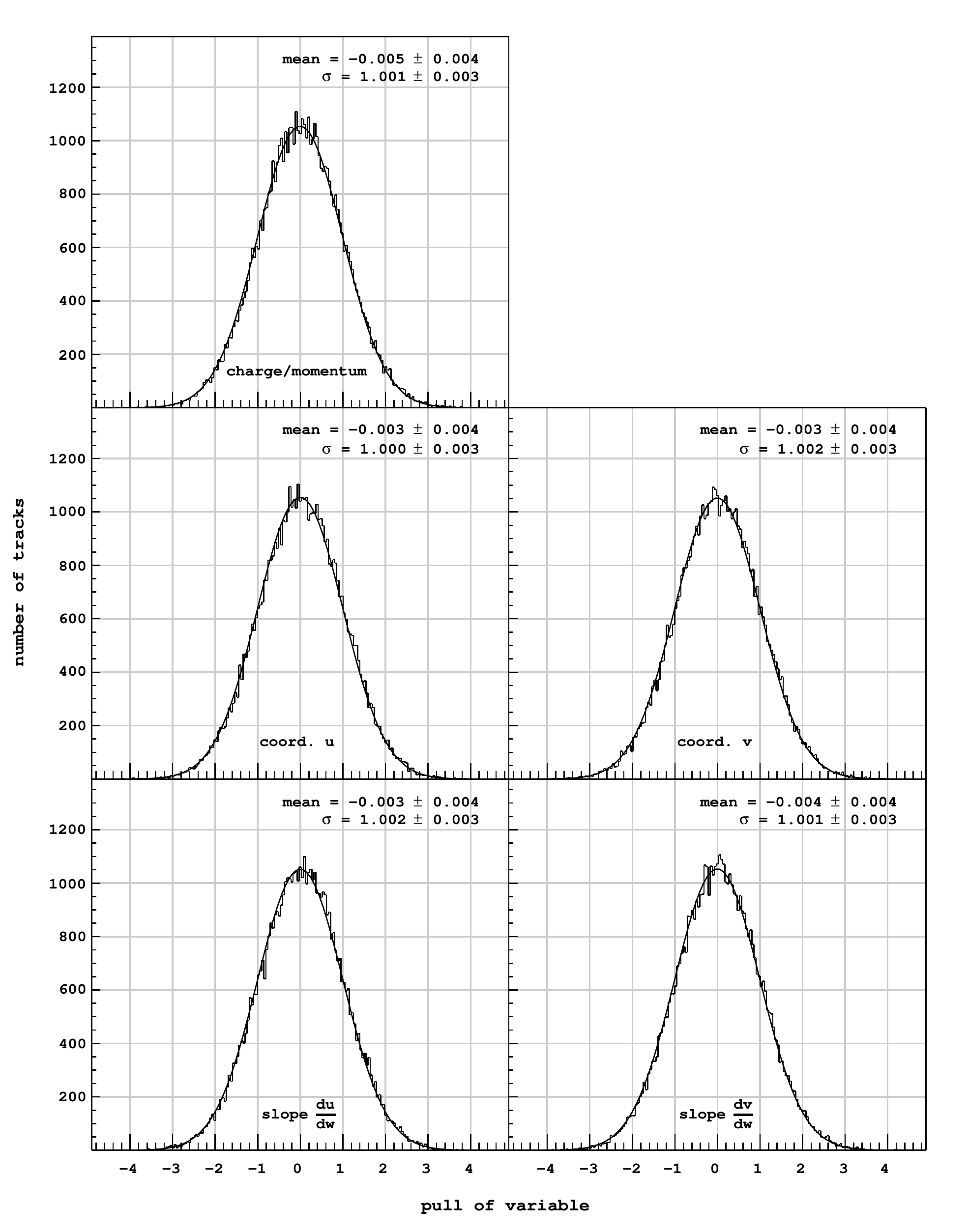}
    \caption {Pull distributions for the five track parameters of
      \code{GeaneTrackRep2}. The pull of a variable $x$ is defined as
    $(x_{\text{fit}}-x_{\text{true}})/\sigma_{x}$.}
    \label{fig:genfit:pulls}
  \end{center}
\end{figure*}

\begin{figure}[htb]
  \begin{center}
    \includegraphics[width=.99\columnwidth]{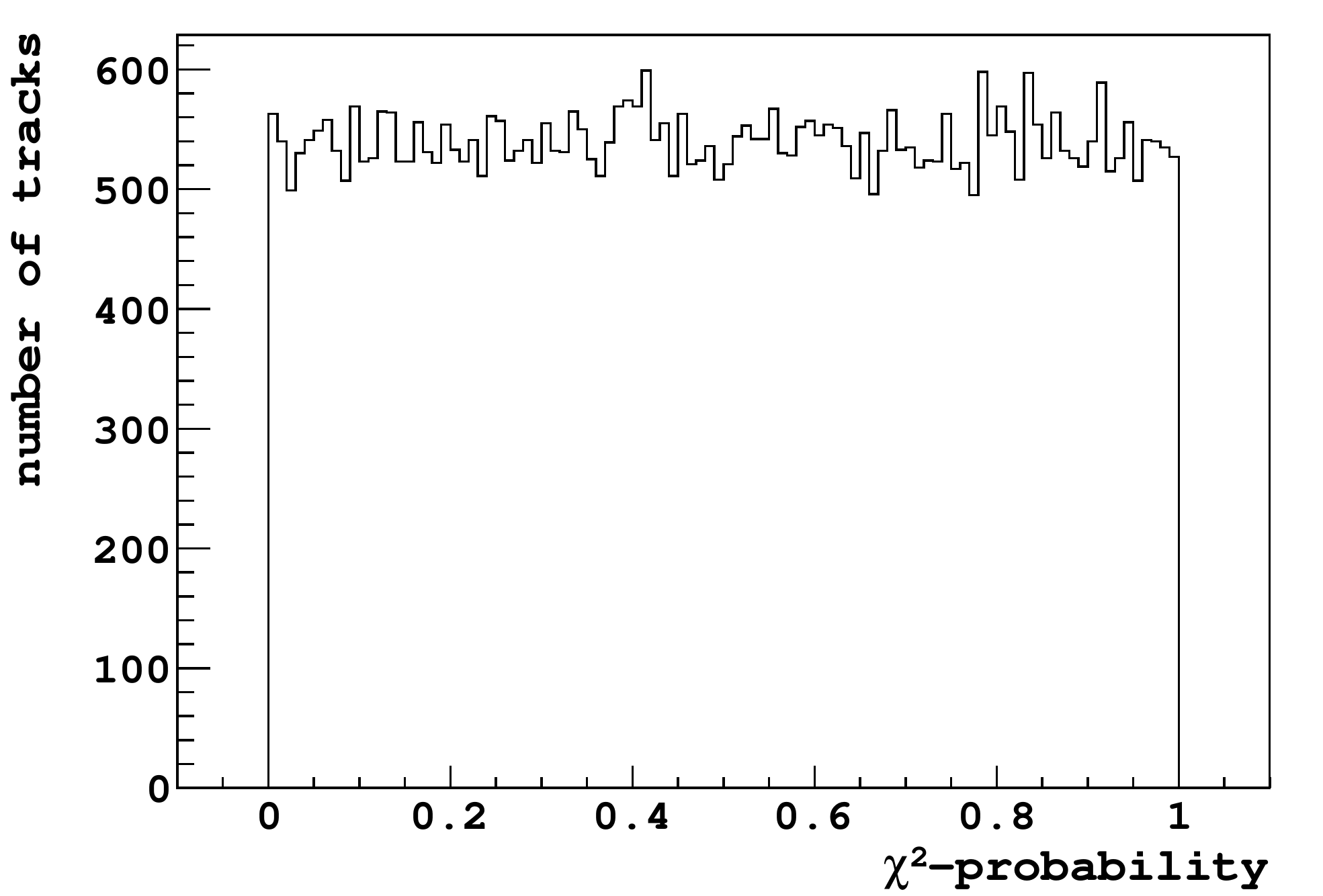}
    \caption{$\chi^2$-probability distribution for a series of track
      fits through 15 planes of strip detectors and 15 space
      points.}
    \label{fig:genfit:pval}
  \end{center}
\end{figure}

The statistical and numerical correctness of a Kalman filter fit
depends on the following items:
  1) The mathematics of the Kalman filter have to be implemented
  correctly. 2) The projections of the covariance matrices of the hits
  onto the (virtual) detector planes have to be correct. 3) The propagation of the track
  parameters and the covariance matrix are done correctly. For the
  covariance matrix this means the correct estimation of the Jacobian
  matrices needed for the Gaussian error propagation. 4) The effects
  of traversed materials must be taken into account correctly: the
  state vector has to be modified (momentum loss) and the entries of
  the covariance matrix need to be increased by the addition of noise
  matrices (e.g.\ due to multiple scattering) \cite{KalmanFit}. Since
  the track representations are external modules, the Kalman
  filter and the reconstruction hit implementation in GENFIT are
  tested with a simplified setup, where the particles traverse a
  vacuum. This way, the effects number 1 to 3 are tested, while the
  effect number 4 is decoupled and not tested here. The setup contains
  a homogeneous magnetic field, since possible problems arising from
  field inhomogeneities would only point to problems in the external
  track representation module and not in the GENFIT core classes.
Instead of detector responses with full digitization simulations,
which result in unknown detector resolutions, known measurement errors
are used.\\ The track representation \code{GeaneTrackRep2} is used for
these tests. The program samples 30 space points on the
trajectory at distances of 1\,cm, which are smeared with Gaussian
distributions of known widths. Like in a TPC, the $x$- and
$y$-measurements are assumed to have equal and better resolutions than
the $z$-coordinate measurements ($\sigma_x=\sigma_y=1/2\cdot
\sigma_z$). These smeared points are used in the fit as reconstruction
hits based on \code{GFSpacepointPolicy} similar to TPC measurements
(see Fig.\ \ref{fig:genfit:recohit}). In front of the first hit, a
reference plane is defined in which the fitted track parameters are
compared to their true values to obtain residual and pull
distributions\footnote{the pull of a variable $x$ is defined as
  $(x_{\text{fit}}-x_{\text{true}})/\sigma_x$.}. If the fit is able to
correctly determine the track parameters and their errors, the pull
distributions will be Gaussians of width $\sigma=1$ and of mean value
$0$.  Figure \ref{fig:genfit:pulls} shows the five pull distributions
for the track parameters, which fulfill these criteria within the corresponding
errors, proving that the non-uniform errors of the
hits are taken into account correctly.\\
Another test is carried out with a slightly different detector
geometry. Hits from 15 crossed planes of strip detectors are fitted
together with 15 space-point hits. The strip hits each contribute one
degree of freedom, the space-point hits each contribute two degrees of
freedom (they only constrain the track in a plane perpendicular to the
track), and the track parameters subtract five degrees of freedom
($15+2\cdot 15-5=40$). The $\chi^2$-probabilities for these fits are
shown in Fig.\ \ref{fig:genfit:pval}. If the number of degrees of
freedom is taken into account correctly, this distribution is expected
to be flat. A $\chi^2$-test against a uniform distribution results in a
$\chi^2 / \text{n.d.f.} = 87.1/99$, close to unity, as expected.\\
The execution time per
  track is 14\,ms on one core of an AMD
  Phenom\textsuperscript{\texttrademark} II X4 940 CPU for 30
  space-point hits with one forward and one backward fitting pass of
  all hits. Of this
  time, a fraction of 91\% is spent in the external extrapolation routines of
  \code{GeaneTrackRep2}, as determined with Valgrind \cite{VALGRIND}.
The GENFIT core classes have not yet been optimized for execution time, but the above
result shows that optimizations would be most rewarding in the track
extrapolation routines.
%\begin{figure}[htb]
 % \begin{center}
  %  \includegraphics[width=.99\columnwidth]{chi2.pdf}
  %  \caption{Reduced $\chi^2$-distribution for a fitting
  %    benchmark. The fit was carried out with 15 planes of strip
  %    detectors and 15 space points. The black line is the fitted
  %    analytical
  %    reduced $\chi^2$-distribution.}
   % \label{fig:genfit:chi2}
  %\end{center}
%\end{figure}

\section{Conclusions and Outlook}
\label{sec:conclusions}
A novel framework for track fitting in particle physics experiments
has been presented in this paper. Its implementation is a C++ library
called GENFIT, which is available freely.  Its modular design consists
of three major building blocks: Fitting algorithms, track
representations, and reconstruction hits. GENFIT contains a
  validated Kalman filter. A standard Kalman smoother is planned to be
  implemented in the future, as well as other fitting algorithms. The
  possibility of the application of GENFIT to pattern recognition
  tasks seems promising and will be
  investigated.\\
The generic design of the track representation interface enables the
user to use any external track following code with GENFIT. The
framework allows simultaneous fits of the same particle track with
different track representations. Possible applications of this feature
are the fitting of different mass hypotheses with the same track
model, or the test and validation of different track parameterizations
and track following codes. Also the coverage of different regions of
phase space with specialized track representations is an important
feature in many experiments. At present, GENFIT contains two track
representations which provide interfaces to the track following code
GEANE and a Runge-Kutta based track extrapolation code adopted from
the COMPASS experiment. New track representations which allow the use
of other track following codes can be implemented in a straightforward
way. The interfaces to the detector geometry and the magnetic field
maps can be chosen freely and are all encapsulated in the track
representation class.\\ The geometrical properties of reconstruction
hits are not restricted in this framework. The dimensionality of hits
is not fixed to particular values, and the orientation of detector planes
can be chosen freely. Hits from detectors which do not measure the
passage of particles in predefined planes, such as drift chambers or
TPCs, are handled in the concept of virtual detector planes. This
leads to a direct minimization of the perpendicular distances between
the particle tracks and the position measurements from the detectors,
i.e.\ the surfaces of
constant drift time or the space points measured in a TPC.\\
GENFIT provides an easy-to-use toolkit for track fitting to the
community of nuclear and particle physics. It is used in the
  PANDA computing framework. Applications in other experiments are
  being considered (e.g.\ Belle II).

\section*{Acknowledgements}
\label{sec:acknowledgements}
This project has been supported by the Sixth Framework Program of the EU
(contract No. RII3-CT-2004-506078, I3 Hadron Physics) and the German
Bundesministerium f\"ur Bildung und Forschung.

%% The Appendices part is started with the command \appendix;
%% appendix sections are then done as normal sections
%\appendix

%% \section{}
%% \label{}
%\bibliographystyle{elsarticle-num}
%\bibliography{genfit}

\begin{thebibliography}{10}
\expandafter\ifx\csname url\endcsname\relax
  \def\url#1{\texttt{#1}}\fi
\expandafter\ifx\csname urlprefix\endcsname\relax\def\urlprefix{URL }\fi
\expandafter\ifx\csname href\endcsname\relax
  \def\href#1#2{#2} \def\path#1{#1}\fi



\bibitem{GEANE}
M.~{Innocente}, V.~{Mairie}, E.~{Nagy}, {GEANE: Average Tracking and Error
  Propagation Package}, CERN Program Library, W5013-E (1991).

\bibitem{VMC}
I.~{Hrivnacova}, D.~{Adamova}, V.~{Berejnoi}, R.~{Brun}, F.~{Carminati},
  A.~{Fasso}, E.~{Futo}, A.~{Gheata}, I.~{Gonzalez Caballero}, A.~{Morsch},
  {for the ALICE Collaboration}, {The Virtual Monte Carlo}, ArXiv Computer
  Science e-prints, cs/0306005.

\bibitem{RECPACK}
A.~{Cervera-Villanueva}, J.~J. {Gomez-Cadenas}, J.~A. {Hernando}, {''RecPack''
  a reconstruction toolkit}, Nuclear Instruments and Methods in Physics
  Research A 534 (2004) 180--183.


\bibitem{PANDA}
{The PANDA Collaboration}, {Physics Performance Report for PANDA: Strong
  Interaction Studies with Antiprotons}, ArXiv e-prints 0903.3905.

\bibitem{PANDAroot}
S.~Spataro, \href{http://stacks.iop.org/1742-6596/119/032035}{Simulation and
  event reconstruction inside the pandaroot framework}, Journal of Physics:
  Conference Series 119~(3) (2008) 032035 (10pp).
%\newline\urlprefix\url{http://stacks.iop.org/1742-6596/119/032035}

\bibitem{genfitSF}
http://sourceforge.net/projects/genfit.


\bibitem{KalmanFit}
R.~{Fr{\"u}hwirth}, {Application of Kalman filtering to track and vertex
  fitting}, Nuclear Instruments and Methods in Physics Research A 262 (1987)
  444--450.

\bibitem{gaussSum}
G.~Kitagawa, The two-filter formula for smoothing and an implementation of the
  gaussian-sum smoother, Annals of the Institute of Statistical Mathematics
  46~(4) (1994) 605--623.

\bibitem{daf}
R.~{Fr{\"u}hwirth}, A.~{Strandlie}, {Track fitting with ambiguities and noise:
  a study of elastic tracking and nonlinear filters}, Computer Physics
  Communications 120 (1999) 197--214.

\bibitem{Kalman}
R.~E. Kalman, A new approach to linear filtering and prediction problems,
  Transactions of the ASME--Journal of Basic Engineering, Series D 82 (1960)
  35--45.

\bibitem{Stroustrup}
B.~Stroustrup, {The C++ Programming Language, Special Edition}, Addison-Wesley
  Longman, Amsterdam, 2000.

\bibitem{ROOT}
R.~Brun, F.~Rademakers, Root - an object oriented data analysis framework, in:
  AIHENP'96 Workshop, Lausane, Vol. 389, 1996, pp. 81--86.


\bibitem{alexandrescu}
A.~Alexandrescu, Modern C++ Design: Generic Programming and Design Patterns
  Applied, Addison-Wesley Professional, 2001.


\bibitem{COMPASS}
{The COMPASS Collaboration}, {The COMPASS experiment at CERN}, Nuclear
  Instruments and Methods in Physics Research A 577 (2007) 455--518.


\bibitem{VALGRIND}
http://valgrind.org.


\end{thebibliography}

%\begin{thebibliography}{00}

%% \bibitem{label}
%% Text of bibliographic item

%\bibitem{}
%\end{thebibliography}
\end{document}